# A pot-pourri of results in QCD from large lattice simulations on the CM5*

Tanmoy Bhattacharya and Rajan Gupta [a]

[a]T-8 Group, MS B285, Los Alamos National Laboratory, Los Alamos, New Mexico 87545 U. S. A.

We present a status report on simulations being done on $32^3 \times 64$ lattices at $\beta = 6.0$ using quenched Wilson fermions. Results for the spectrum, decay constants, the kaon B-parameter $B_K$, and semi-leptonic form factors are given based on the current statistical sample of 28 configurations. These "Grand Challenge" calculations are being done on the CM5 at Los Alamos in collaboration with G. Kilcup, S. Sharpe and P. Tamayo. We end with a brief statement of the performance of our SIMD code on the CM5.

## 1. LATTICE PARAMETERS

All the results described in these two talks have been obtained using the following lattice parameters. The $32^4$ gauge lattices were generated at $\beta = 6.0$ using the combination 10 over-relaxed (OR) sweeps followed by 1 Metropolis sweep. We have stored lattices every 2500 OR sweeps. Quark propagators, using the simple Wilson action, have been calculated after doubling the lattices in the time direction, *i.e.* ($32^3 \times 32 \rightarrow 32^3 \times 64$). This doubling was done in response to the initial hardware/operational constraints of the CM5 at LANL. Currently we are generating $32^3 \times 64$. Periodic boundary conditions are used in all 4 directions, both during lattice update and propagator calculation. Quark propagators have been calculated using two kinds of extended sources – Wuppertal and Wall – at $\kappa = 0.135$ ($C$), 0.153 ($S$), 0.155 ($U_1$), 0.1558 ($U_2$), and 0.1563 ($U_3$). These quark masses correspond to pseudoscalar mesons of mass 2800, 980, 700, 550 and 440 $MeV$ respectively where we have used $1/a = 2.3GeV$ for the lattice scale. The three light quarks allow us to extrapolate the data to the physical $u$ and $d$ quarks, while the $C$ and $S$ $\kappa$ values are selected to be close to the physical charm and strange quark masses. The $C$ ($S$) quark mass is slightly lower (higher) than the physical value. So far our statistical sample consists of 28 configurations, each separated by 5000 OR sweeps. The goal is to analyze 100 configurations so the present analysis should be considered preliminary. Also, Sharpe [4] and Bernard and Golterman [3] have pointed out that in the quenched approximation $\eta'$ loops give rise to unphysical terms in the chiral expansion for observables. These effects have not been included in the analysis when extrapolating quantities to the chiral limit.

We find that the data for almost all observables show a significant curvature as a function of the quark mass. So we make both a linear and a quadratic fit in each case and quote final numbers from the fit with lower $\chi^2$. More details about the fits are given in the individual sections.

## 2. SPECTRUM

We have analyzed three types of hadron correlators distinguished by the type of source/sink used to generate quark propagators. These are (i) wall source and point sink (WL), (ii) Wuppertal source and point sink (SL), and Wuppertal source and sink (SS). With these three cases we are able to make a consistency check since the effective mass $m_{eff}(t)$ converges to the asymptotic value from below for WL correlators and from above for SL and SS correlators in all hadron channels. Fig. 1 illustrates, using the pion $U_1U_1$ channel, noteworthy features of the convergence of $m_{eff}(t)$. In the case of WL correlators, the approach of $m_{eff}(t)$ to its asymptotic value is monotonic but quite slow and one needs to fit for $t > 15$. The statistical errors in $m_{eff}(t)$ are smallest of the three cases. The SS correlators reach the asymptotic value at $t \approx 8$ and the time-slice to time-slice fluctuations are of the order of

*Talks presented by T. Bhattacharya and R. Gupta.

Figure 1. Comparison of the convergence of $m_{eff}(t)$ for three types of pion $(U_1U_1)$ correlators. $SL$ data has been shifted by $+0.05$.

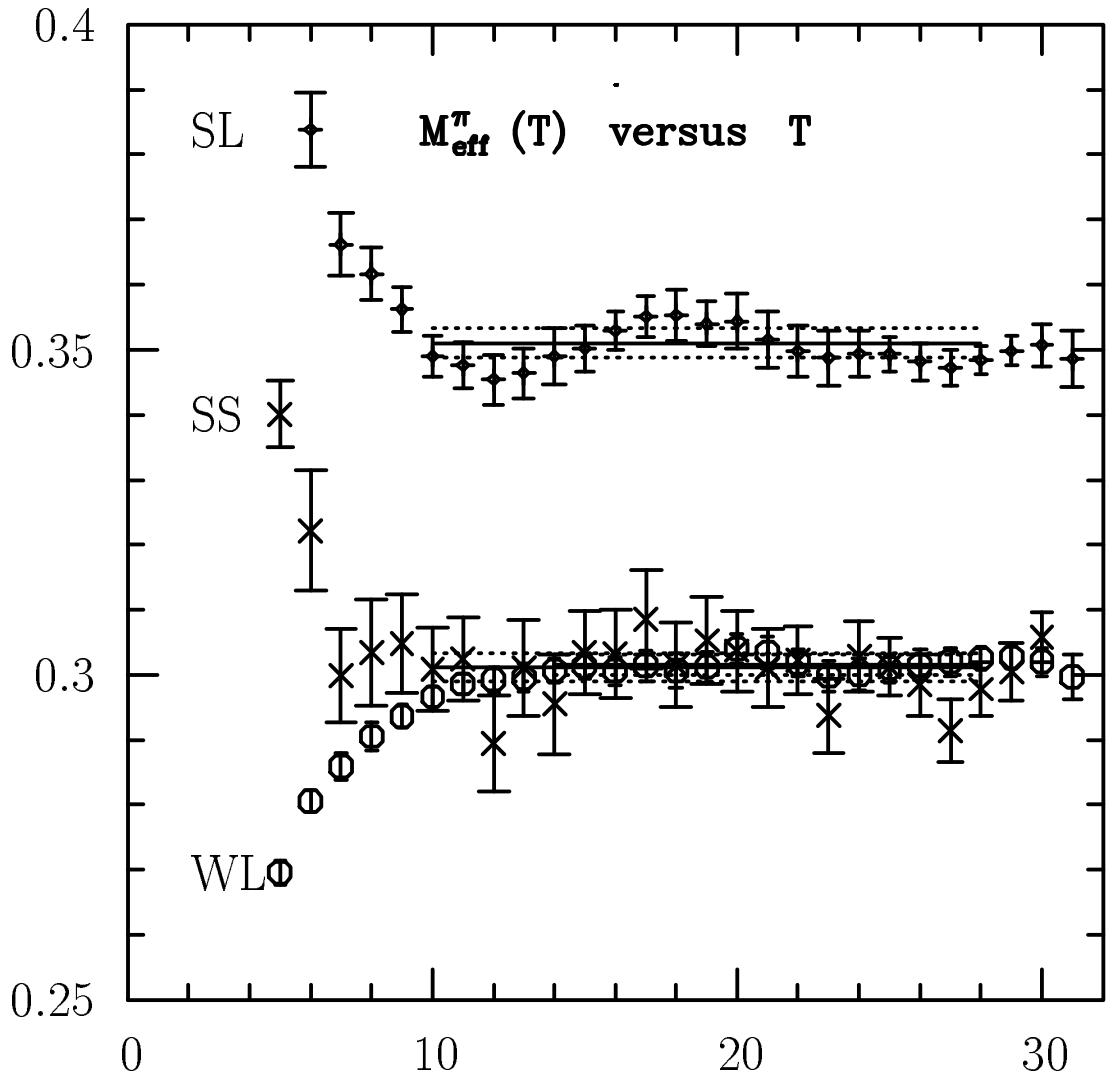

Figure 2. Quadratic extrapolation of $\Delta$, nucleon and $\rho$ mass data to the chiral limit.

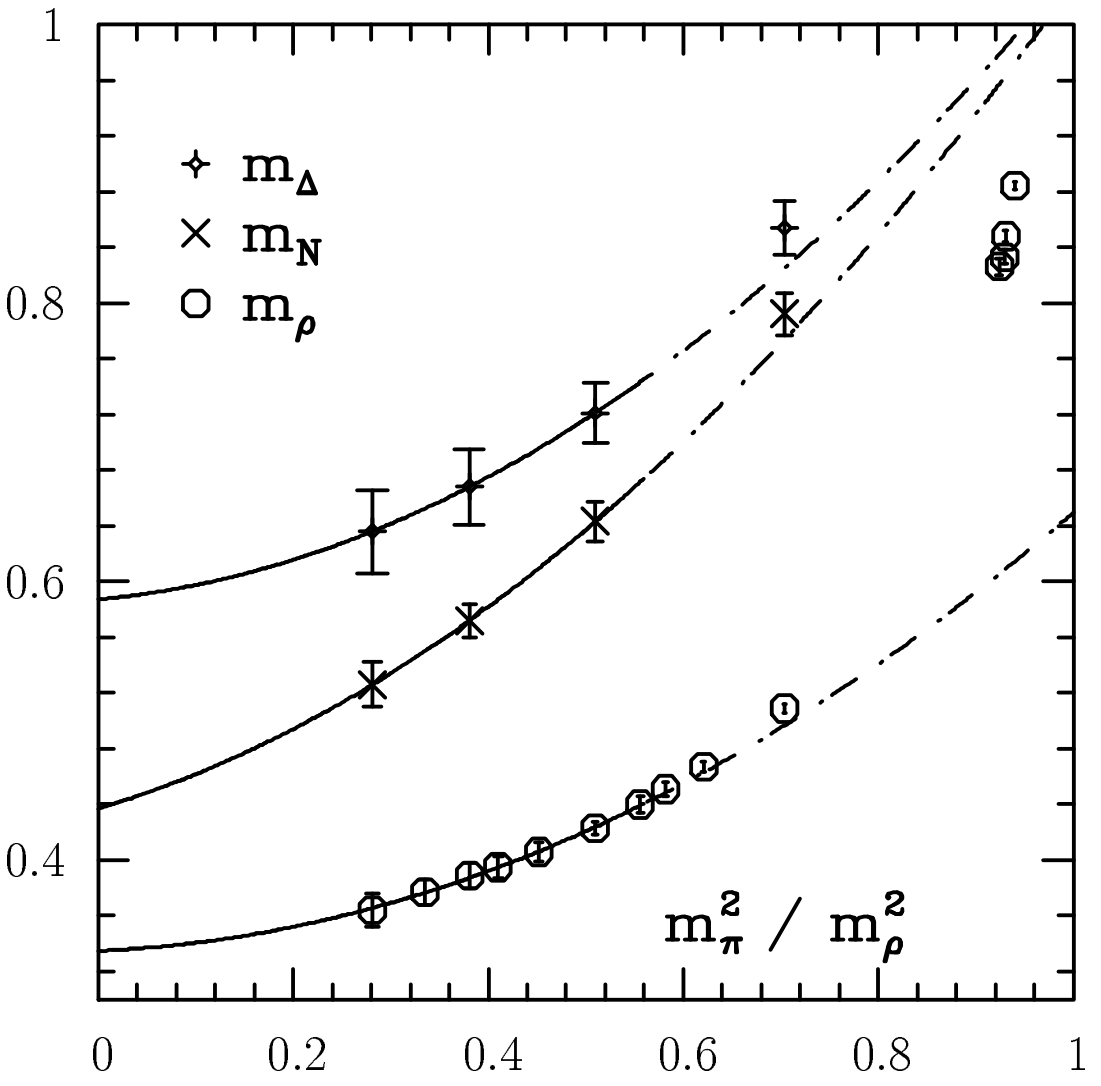

the statistical errors. The SL correlators show correlated fluctuations in $m_{eff}(t)$, wiggles with a period of about 12 time slices, which may be a feature of the Wuppertal source. Another possibility, which we will check on the new set of $32^3 \times 64$ lattices, is that this feature is due to doubling the lattice in the time direction. Inspite of these differences in $m_{eff}(t)$, the bottom line is that all three kinds of correlators give consistent (within 1 $\sigma$) results for masses for all mesons and the error estimates are similar too. We give the average values for masses in lattice units in Table 1.

To extrapolate the data to the chiral limit in the light quark we fit as a function of $m_\pi^2/m_\rho^2$ and $m_q$, where the latter is defined either as $\log(1+0.5(1/\kappa - 1/\kappa_c))$ or $0.5(1/\kappa - 1/\kappa_c)$, or determined non-perturbatively as discussed in Ref. [1]. Using a quadratic extrapolation of $m_\pi^2$ data versus $m_q$ (as it fits all the points) we get

$$\kappa_c = 0.15721(7). \qquad (1)$$

This value is higher that given in Ref. [1], but consistent with QCDPAX result 0.15717(3). Similarly, using a quadratic fit, the extrapolated value of $m_\rho$ is (see Fig. 2)

$$m_\rho(m_q = 0) = 0.34(4) \qquad (2)$$

which gives a lattice scale of

$$a^{-1}(m_\rho) = 2.26(27)\ GeV. \qquad (3)$$

We shall, for simplicity, use $1/a = 2.3\ GeV$.

Wuppertal source quark propagators allow one to construct hadron correlators with non-zero momenta. In Fig. 3 we show the pion propagator at 4 different momenta. The signal in $\vec{p} = (1,0,0)$ and $(1,1,0)$ correlators is quite good. Thus, as we show later, using $32^3$ lattices at $\beta = 6.0$ leads to a significantly improved signal in the calculation of phenomenologically interesting quantities like semi-leptonic form factors and $B_K$.

### 2.1. Baryons

In our previous work[1] we had found that wall and Wuppertal sources give results that are significantly different. In Fig. 4 we show a typical example of the current status. We find that on the larger lattices $WL$ and $SL$ correlators give consistent results. The plateau in $m_{eff}(t)$ is not very clean and any systematic difference between the two cases is an artifact of the fit range and/or the statistical errors. The numbers given in Table 1 for the case of degenerate quarks are the mean of $SL$ and $WL$ results. Extrapolating the

Figure 3. Effective mass plot for pion ($U_1U_1$) at different momenta.

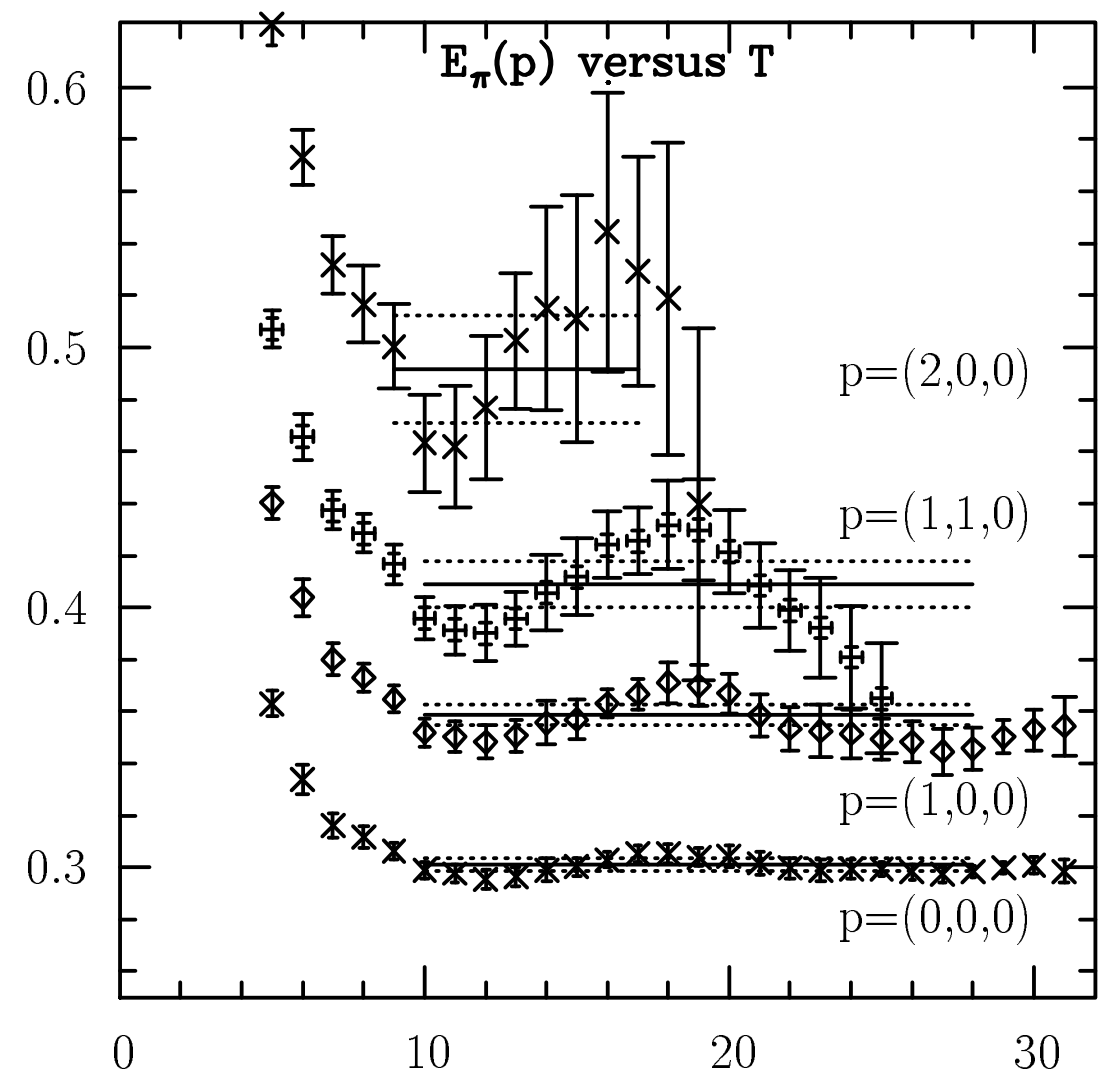

data to the chiral limit (Fig. 2) gives

$$m_N a = 0.44(20)$$
$$m_\Delta a = 0.59(40)\ , \qquad (4)$$

however, before comparing these numbers to experimental data the reader should note that we have not included the results of Labrenz and Sharpe[6] who predict additional non-analytic terms in the chiral behavior of quenched QCD.

## 3. DECAY CONSTANTS

Meson decay constants can be calculated using $SL$ and $SS$ correlators. We use the generic notation $f_{PS}$ for pseudoscalar mesons and $f_V$ for vector meson. There are many different ways of combining these correlators to extract $f_{PS}$ and $f_V$, some of which are described in Ref. [1]. We find that all methods give essentially identical results, so we quote the mean value in Table 2. For the error estimate we give the largest individual error. We also find that the value of $f_{PS}$ is independent of the pion's momentum. For comparison we give results using the local current with two different renormalization factors. Columns 2 and 4 are using the Lepage-Mackenzie mean-field improved scheme[5] while columns 1 and 3 use the naive $\sqrt{2\kappa}$ renormalization with $Z_v = 0.57$ and the 1-loop perturbative value for $Z_A$. In both cases we use a boosted value for the coupling, $g^2 = 1.75$. The results are displayed in Fig. 5.

As shown in Fig. 5, data for $f_\rho^{-1}$ is qualitatively different for cases with degenerate versus non-degenerate quark masses. The large dependence on the renormalization constant (naive versus improved) even for small $m_q$ suggests possibly large $O(a)$ corrections or effects of quenching in $f_\rho^{-1}$. To get $f_\pi$ we can extrapolate the degenerate combinations or all data with $S$ and lighter quarks as they give qualitatively similar results.

Table 1
Masses in lattice units

| | $m_\pi$ | $m_\rho$ | Nucleon | $\Delta$ |
|---|---|---|---|---|
| $CC$ | 1.219(2) | 1.232(02) | | |
| $CS$ | 0.857(2) | 0.884(03) | | |
| $CU_1$ | 0.818(3) | 0.848(04) | | |
| $CU_2$ | 0.803(4) | 0.833(05) | | |
| $CU_3$ | 0.794(6) | 0.826(06) | | |
| $SS$ | 0.427(2) | 0.509(03) | 0.792(15) | 0.854(19) |
| $SU_1$ | 0.368(2) | 0.467(04) | | |
| $SU_2$ | 0.344(2) | 0.451(05) | | |
| $SU_3$ | 0.328(2) | 0.440(06) | | |
| $U_1U_1$ | 0.302(2) | 0.423(05) | 0.643(14) | 0.721(22) |
| $U_2U_2$ | 0.240(2) | 0.389(09) | 0.572(12) | 0.668(27) |
| $U_3U_3$ | 0.193(2) | 0.364(12) | 0.526(16) | 0.636(30) |

Table 2
$f_\pi$ and $1/f_V$ in lattice units

| | $f_\pi^{naive}$ | $f_\pi^{imp.}$ | $1/f_V^{naive}$ | $1/f_V^{imp.}$ |
|---|---|---|---|---|
| $CC$ | 0.126(3) | 0.205(6) | 0.097(2) | 0.20(1) |
| $CS$ | 0.103(3) | 0.137(4) | 0.121(3) | 0.20(1) |
| $CU_1$ | 0.094(3) | 0.122(4) | 0.116(3) | 0.19(1) |
| $CU_2$ | 0.090(3) | 0.116(4) | 0.115(4) | 0.19(1) |
| $CU_3$ | 0.088(4) | 0.113(5) | 0.114(4) | 0.18(1) |
| $SS$ | 0.090(3) | 0.098(3) | 0.228(6) | 0.31(1) |
| $SU_1$ | 0.083(3) | 0.089(3) | 0.241(7) | 0.32(1) |
| $SU_2$ | 0.080(3) | 0.084(3) | 0.246(7) | 0.33(1) |
| $SU_3$ | 0.078(3) | 0.081(3) | 0.248(8) | 0.33(1) |
| $U_1U_1$ | 0.077(3) | 0.080(3) | 0.261(7) | 0.34(1) |
| $U_2U_2$ | 0.071(3) | 0.072(3) | 0.276(8) | 0.35(1) |
| $U_3U_3$ | 0.066(4) | 0.067(4) | 0.285(9) | 0.36(1) |

Figure 4. Effective mass plot for nucleon with $WL$ and $SL$ correlators.

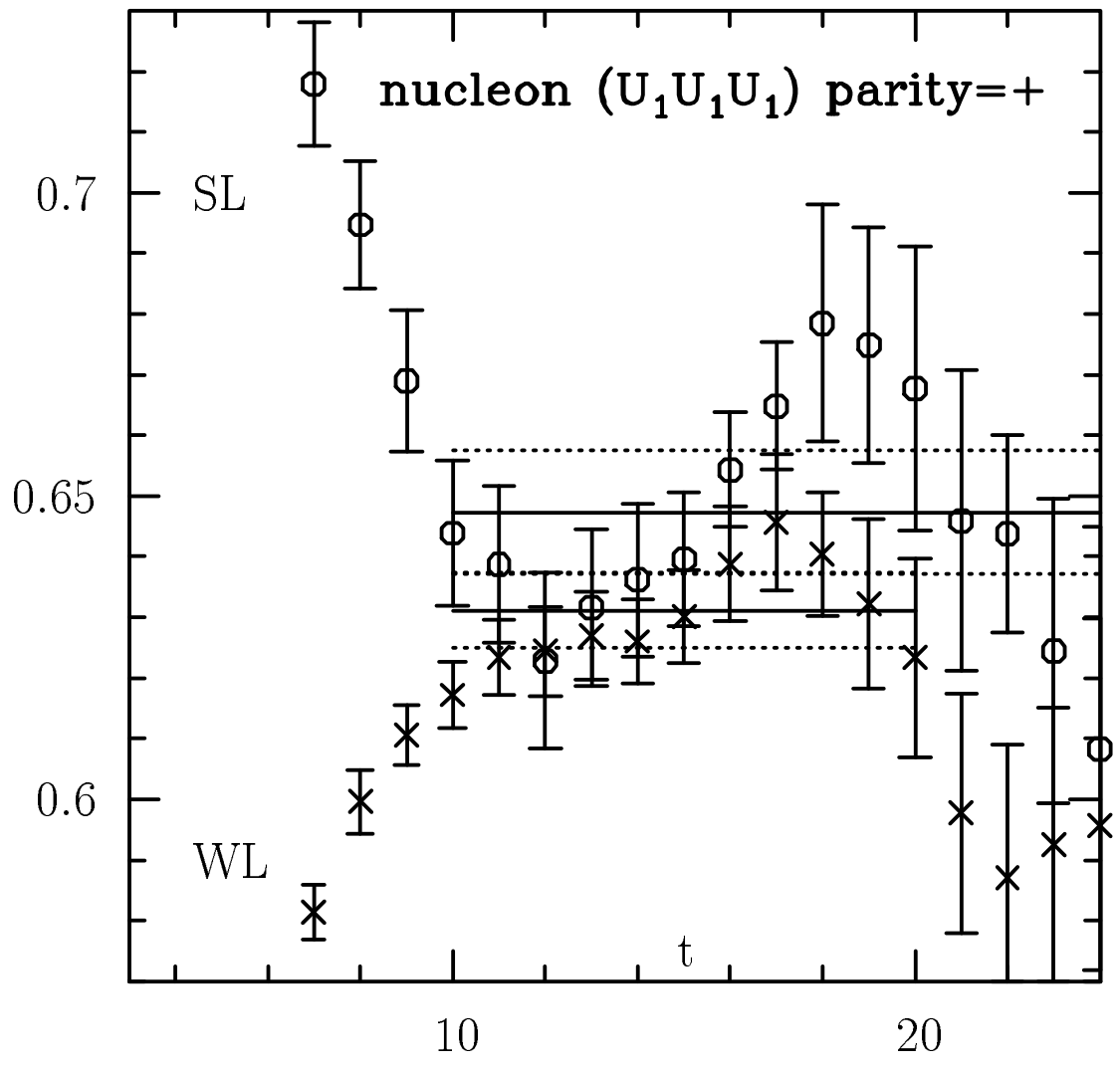

Using the mean-field improved normalizations we get (with all data)

$$f_\pi a = 0.052(5) \qquad \text{linear}$$
$$f_\pi a = 0.057(24) \qquad \text{quadratic} \,. \qquad (5)$$

Both fits have comparable $\chi^2$. From the linear fit we get $1/a = 2.5(3)GeV$. For $f_D$ we calculate the ratio $f_D/f_K$ using $S$ $(U_1)$ as the strange quark and get the result 1.38 (1.52), where the change due to the light quark mass is $\approx 1$ in the last decimal place. Using $f_K = 160MeV$ and the mean of the above two numbers ($\approx$ physical $s$ quark) we get $f_D = 232(20)MeV$.

## 4. $B_K$

The major problem in the calculation of the kaon $B$-parameter with Wilson fermions is the bad chiral behavior induced by the mixing of the $\Delta S = 2$ 4-fermion operator with wrong chirality operators. This mixing arises as a result of the $r$-term in the action and has been calculated to 1-loop in perturbation theory. In Ref. [2] it was shown that the lattice artifacts completely overwhelm the signal when using the 1-loop improved operator, however, by calculating the matrix element at different values of momentum transfer one can remove most of these. The main limitation of that calculation was that the spatial lattice size was small ($16^3$), consequently the momentum gap ($2\pi/16$) was large, and the signal in the non-zero momentum correlators was marginal. Both these limitations have been addressed on the present set of lattices and as a result the statistical quality of the data is markedly improved.

Figure 5. Meson decay constants $f_{PS}$ and $f_V^-1$ as a function of $m_\pi^2/m_\rho^2$. Data for $f_V^{-1}$ has been shifted by +0.25 for clarity.

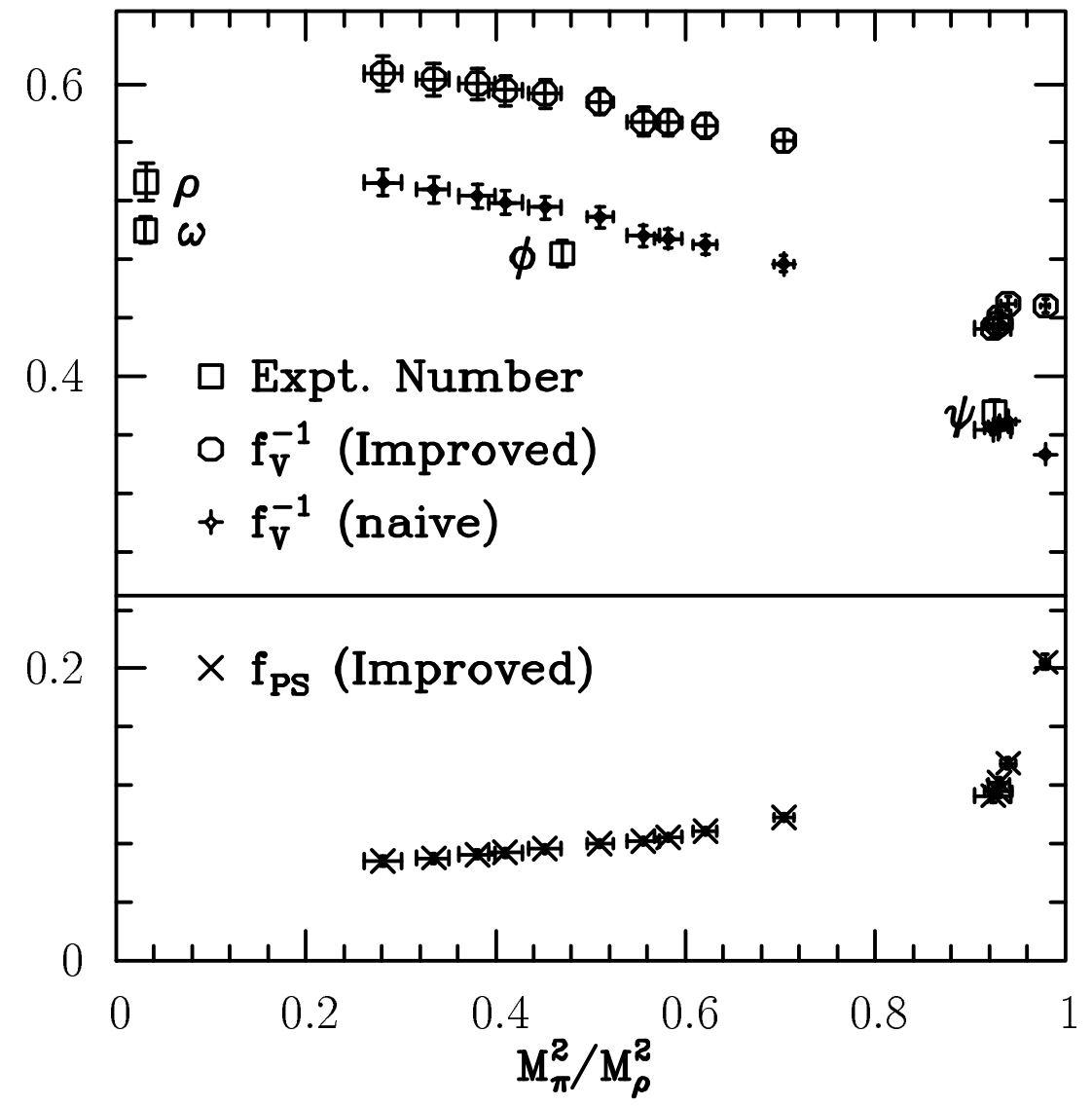

In Fig. 6 we show the behavior of the lattice $B_K$, evaluated at momentum transfer $p = 2\pi/32$, as a function of the time slice at which the operator is inserted. The data shown are for two different kaon sources, *i.e.* $\bar\psi_s\gamma_5\psi_d$ and $\bar\psi_s\gamma_4\gamma_5\psi_d$ as they converge from opposite direction. The statistical errors are small and the two fits agree within errors. The results for $p = 2\pi/32 \times (1,1,0)$ and $p = 2\pi/32 \times (1,1,1)$ are also reliable. Full analysis will be presented elsewhere Ref. [7].

In Table 3 we present preliminary results for the $B$-parameters using perturbatively improved operators evaluated with a boosted $g^2 = 1.75$ and $\mu a = 1.0$. In column 1 (2) we give the raw number for zero (one) unit of momentum transfer. The final values of $B_K$ after removing the leading bad chiral contributions by the method of momentum subtraction (Ref. [2]) are given in column 3. In columns 4 and 5 we give the results for the two left-right operators $O_7$ and $O_8$ (defined in Ref. [2]).

Figure 6. $B_K$ at momentum transfer $\vec{p} = (1,0,0)$ as a function of $t$ at which the operator is inserted. The two data sets correspond to the two kinds of operators used to produce the kaons.

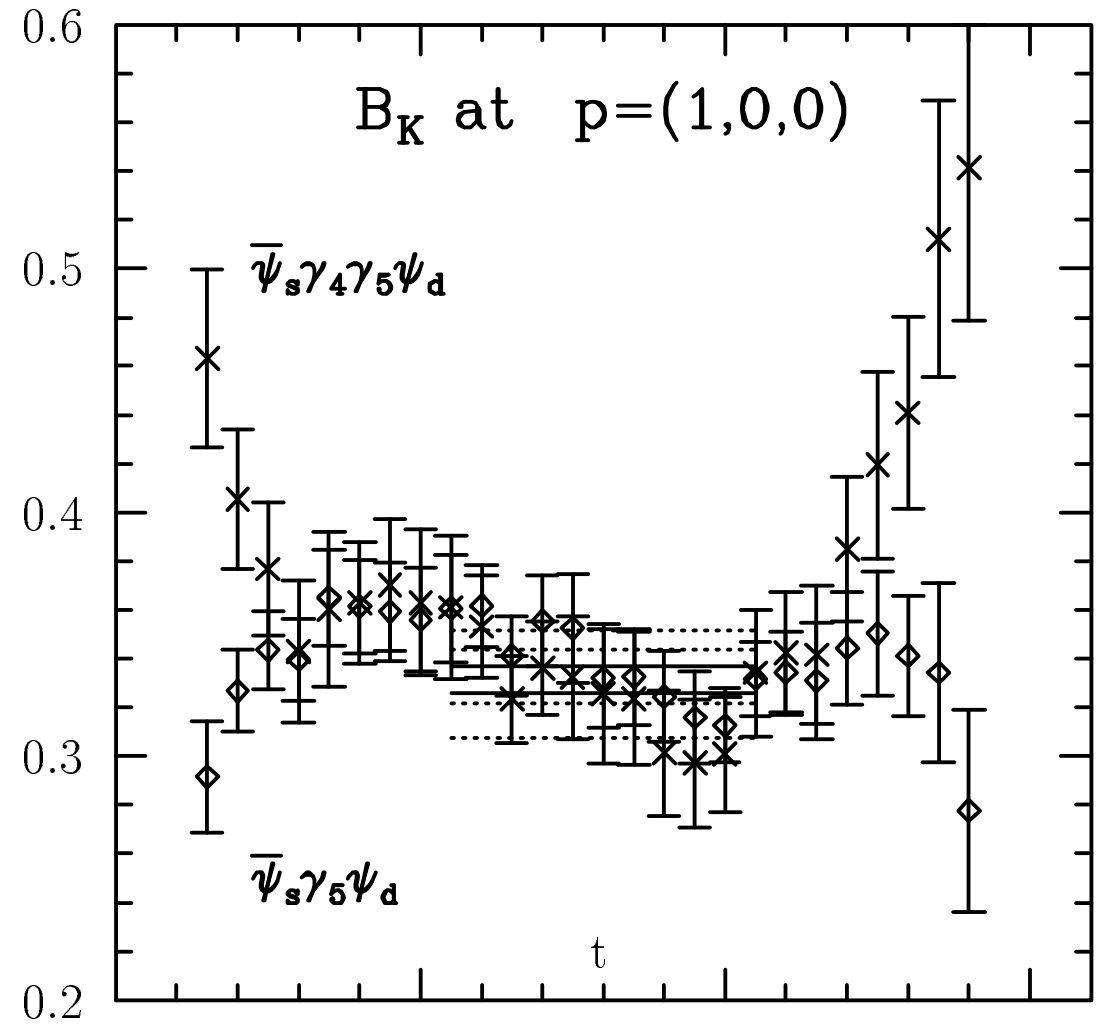

The data in columns 1 and 2 show that the statistical errors have been reduced to the level of a few percent. The reason that the errors in the final $B$-parameters are large for heavy quarks is mainly because $E(p) \approx m$, while for light quarks the statistical fluctuations are large.

The numbers extrapolated to the kaon mass are given in the last two rows in Table 3. We use two methods for this: in the first we keep $S$ mass fixed and extrapolate in the light quark mass using $S$, $U_1$, $U_2$, $U_3$. In the second, the two quarks are taken to be degenerate. Both methods give similar results for all three operators. This suggests that the SU(3) breaking does not give a large contribution to these $B$-parameters. Also, there is little evidence of unphysical contributions from $\eta'$ loops which are expected to be significant near the chiral limit in the quenched approximation[8].

The values of $B_K$, $B_7$ and $B_8$ are significantly lower than those found in Ref. [2]. We believe that this is due to using a larger lattice and a smaller $\vec{p}$. Very accurate and reliable results for $B_K$, obtained using staggered fermions, have been presented at this conference by S. Sharpe[8]. Wilson results are consistent with these but have larger systematic errors as a result of mixing with operators having the wrong chiral behavior. By improving the quality of Wilson fermion results one can, by comparison with staggered results, check our understanding of lattice artifacts.

Table 3
$B_K$, $B_7$ and $B_8$

| | $B_K$ | | | $B_7$ | $B_8$ |
|---|---|---|---|---|---|
| | $(p=0)$ | $(p=1)$ | subtr. | | |
| $CC$ | 0.95(3) | 0.95(4) | 1.32(37) | 0.97(3) | 0.97(3) |
| $CS$ | 0.83(2) | 0.84(3) | 1.09(21) | 0.94(3) | 0.95(3) |
| $CU_1$ | 0.79(2) | 0.80(2) | 0.94(19) | 0.92(3) | 0.94(3) |
| $CU_2$ | 0.77(2) | 0.77(2) | 0.76(20) | 0.91(3) | 0.93(3) |
| $CU_3$ | 0.76(2) | 0.75(3) | 0.50(24) | 0.89(3) | 0.92(3) |
| $SS$ | 0.50(1) | 0.53(1) | 0.82( 7) | 0.87(2) | 0.91(2) |
| $SU_1$ | 0.36(1) | 0.41(2) | 0.72( 6) | 0.84(1) | 0.89(1) |
| $SU_2$ | 0.29(1) | 0.34(2) | 0.66( 7) | 0.83(1) | 0.88(1) |
| $SU_3$ | 0.23(1) | 0.28(2) | 0.60( 8) | 0.82(1) | 0.87(1) |
| $U_1U_1$ | 0.13(1) | 0.21(2) | 0.62( 6) | 0.81(1) | 0.86(1) |
| $U_2U_2$ | −.26(2) | −.08(3) | 0.48( 8) | 0.76(1) | 0.82(1) |
| $U_3U_3$ | −.78(3) | −.42(4) | 0.37(11) | 0.72(1) | 0.78(1) |
| Ext 1 | | | 0.48 | 0.80 | 0.85 |
| Ext 2 | | | 0.43 | 0.74 | 0.80 |

## 5. SEMILEPTONIC MESON FORM FACTORS

We have calculated the semileptonic pseudoscalar form factors of the $D$ meson, both for $D \to Ke\nu$ and $D \to \pi e\nu$. In calculating these form factors we have held the $D$-meson at zero spatial momentum, and varied the momentum of the other meson from 0 to $\pi/8$. This provides a large enough range (both positive and negative) in the invariant mass of the leptonic subsystem ($-Q^2$) to test the pole dominance hypotheses. To control systematic effects, we have done the calculation using three different transcriptions (called 'local', 'extended' and 'conserved') of the current. The details of the method are given in Ref. [9] and a complete analysis for the present set of lattices will be presented in Ref. [10].

Table 4
The meson form factors extrapolated to $Q^2 = 0$

| | $f_+^K(0)$ | $f_+^\pi(0)$ | $f_0^K(0)$ | $f_0^\pi(0)$ |
|---|---|---|---|---|
| $U_1$ | 0.833(21) | 0.787(24) | 0.808(14) | 0.735(14) |
| $U_2$ | 0.837(24) | 0.761(32) | 0.808(15) | 0.699(16) |
| $U_3$ | 0.834(27) | 0.737(41) | 0.806(17) | 0.670(18) |

The renormalization of all three currents is carried out using the Lepage-Mackenzie[5] mean field improved perturbation theory. We find that the values obtained using the three currents are very close. (See Fig. 7 for an example of consistency between different currents.) We notice that the statistical errors are very small when the hadrons carry zero or one unit of momentum, and some systematic difference between the local and the non-local currents become visible. For $\vec{p} = (1,1,1)$ and $(2,0,0)$ there is no clear plateau in the data and the analysis is very sensitive to the fit range.

Except when $\vec{p} = (1,1,1)$, the data are consistent (see Fig. 7 for an example) with the hypothesis that the form factor is dominated by the nearest pole with the right quantum numbers ($D_s^*$ and $D^*$ respectively for $f_+^K$ and $f_+^\pi$). Assuming pole dominance we can extract $f(Q^2 = 0)$, and the results are summarized in Table 4. To within statistical errors the numbers are independent of the light quark mass for $f^K$ and there is a slight decrease in $f^\pi$. We expect $\approx 10\%$ change in $f^\pi$ on extrapolating the data to the chiral limit and to the physical values of charm quark mass. Our final numbers for $f_+^K$ are close to the phenomenological value, $f_+^K \approx f_0^K$ and $f_+^K > f_+^\pi$.

## 6. CODE PERFORMANCE ON THE CM5

Our CM5 code uses the SIMD programming environment. All computationally intensive portions of the code have been written in CDPEAC and modularized. Calling these library of low level codes is not a significant overhead due to the performance of the Sparc chip, while having the algorithm control statements in CM Fortran provides flexibility, making the adding/changing of algorithms and physics analysis easy. Our present gauge update code runs at 25 megaflops/node and the Wilson propagator generation sustains 35 megaflops/node. These timings include all I/O and setup overhead. For the present set of physics objectives we need 18000 node-hours to generate and process one configuration.

Figure 7. Comparison of $1/f_0$ obtained using the three vector currents. The data are for the case where the spectator quark is $U_3$. The fits show comparison with pole dominance using the measured lattice value for the scalar mass, $m_{D_s}$.

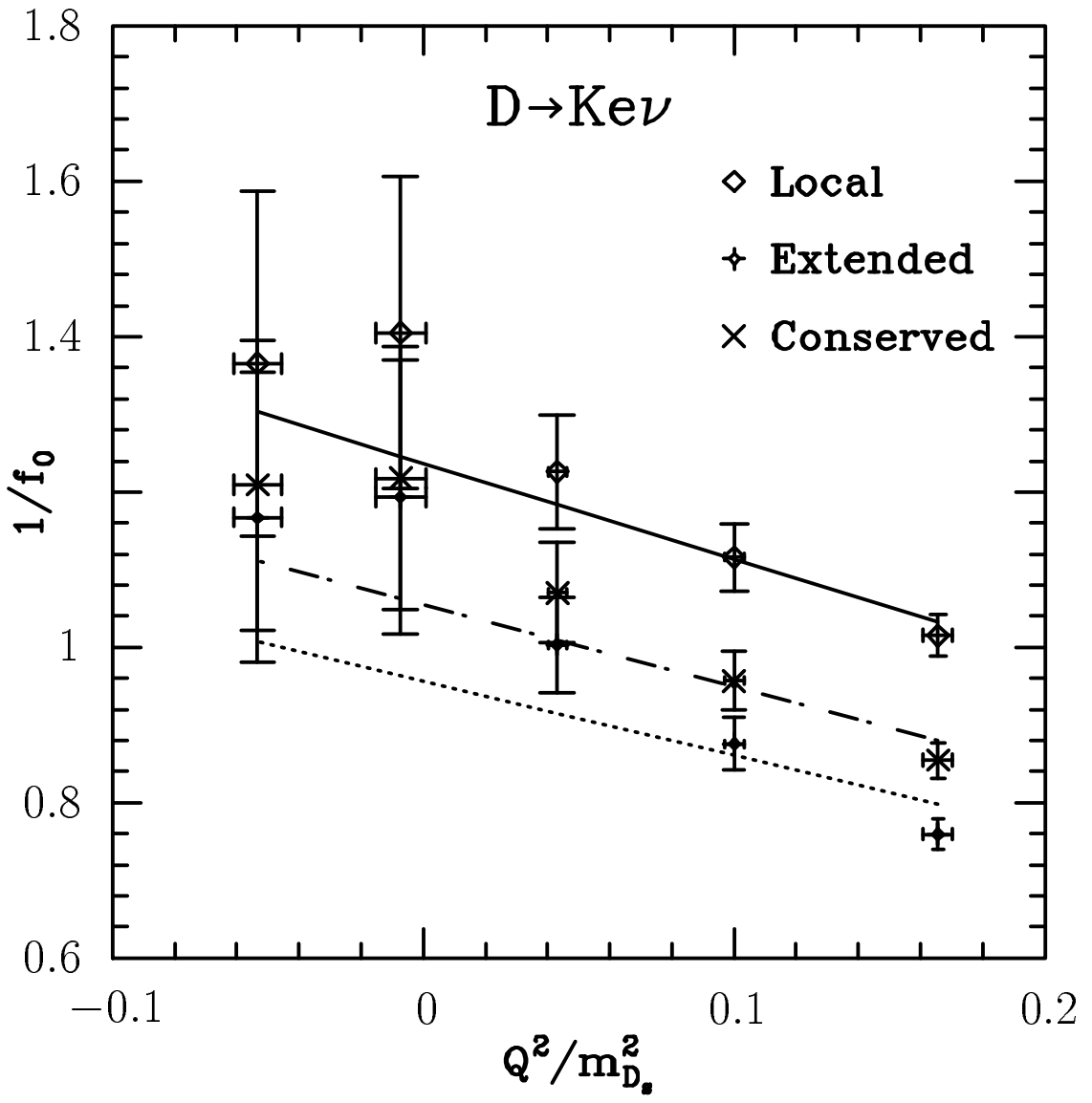

We gratefully acknowledge the tremendous support provided by the ACL at Los Alamos. These calculations have been done as part of the DOE HPCC Grand Challenge program.